\documentclass[twocolumn,prl,lengthcheck,showpacs,showkeys, superscriptaddress, amsmath,amssymb]{revtex4}
\usepackage{graphicx}
\usepackage{bm}
\usepackage{amsmath}
\begin{document}
\preprint{}
\title{Spin waves in paramagnetic BCC iron: spin dynamics simulations}
\author{Xiuping Tao}
\email{txp@uga.edu}
\affiliation{Center for Simulational Physics, University of Georgia, Athens, GA 30602}
\author{D.~P.~Landau}
\affiliation{Center for Simulational Physics, University of
Georgia, Athens, GA 30602}
\author{T.~C.~Schulthess}
\affiliation{Computer Science and Mathematics Division, Oak Ridge
National Laboratory, Oak Ridge, TN 37831-6164}
\author{G.~M.~Stocks}
\affiliation{Metals and Ceramics Division, Oak Ridge National
Laboratory, Oak Ridge, TN 37831-6114}
\date{\today}

\begin{abstract}
Large scale computer simulations are used to elucidate a longstanding 
controversy regarding the existence, or otherwise, of
spin waves in paramagnetic BCC iron. Spin dynamics simulations of
the dynamic structure factor of a Heisenberg model of Fe with
first principles interactions reveal that well defined peaks
persist far above Curie temperature $T_c$. At large wave
vectors these peaks can be ascribed to propagating spin waves, at
small wave vectors the peaks correspond to over-damped spin waves.
Paradoxically, spin wave excitations exist despite only limited magnetic
short-range order at and above $T_c$.
\end{abstract}

\maketitle

For over three decades, the nature of magnetic
excitations in ferromagnetic materials above the Curie temperature
$T_c$ has been a matter of controversy amongst experimentalists
and theorists alike. Early neutron scattering experiments on iron
suggested that spin waves were renormalized to zero at $T_c$
\cite{Collins}; however, in 1975, using unpolarized neutron
scattering techniques, Lynn at Oak Ridge (ORNL) reported
\cite{Lynn1} that spin waves in iron persisted as excitations up
to the highest temperature measured ($1.4T_c$), and no further
renormalization of the dispersion relation was observed above
$T_c$.

Experimentally, it was challenged primarily by
Shirane and collaborators at Brookhaven (BNL) \cite{Shirane1}.
Using polarized neutrons, they reported that spin wave modes were
not present above $T_c$ and suggested that the ORNL group
needed polarized neutrons to subtract the background scattering
properly. Utilizing full polarization analysis techniques the ORNL
group subsequently confirmed their earlier work and, in addition,
they analyzed data from both groups and concluded that their
resolution was more than an order of magnitude better than that
employed by the BNL researchers \cite{Lynn2}. Moreover, angle-resolved
photoemission studies \cite{Haines1, Kisker} suggested
the existence of magnetic short-range order (SRO) in paramagnetic
iron and that this could give rise to propagating modes.  
Theoretically, SRO of rather long length scales (25 \AA) was 
postulated to exist far above $T_c$ \cite{Korenman1, Capellmann1} and
a more subtle kind was proposed later \cite{Heine}. Contrarily, it was
also suggested that above $T_c$, all thermal excitations are
dissipative \cite{Hubbard1, Moriya}. To further complicate
matters, analytical calculations for a Heisenberg model of iron,
with exchange interactions extending to fifth-nearest neighbors
and a three pole approximation \cite{Shastry1}, did not reproduce
the line shape measured by either experimental group mentioned
above. In addition, Shastry \cite{Shastry2} performed spin
dynamics (SD) simulations of a nearest neighbor Heisenberg model
of paramagnetic iron with $8192$ spins and showed some plots of
dynamic structure factor $S({\bf q},\omega )$ with a shoulder at
nonzero $\omega$ for some ${\bf q}$. It was explained to be due to statistical
errors instead of propagating modes.

With new algorithmic and computational capabilities, 
qualitatively more accurate SD simulations can now be performed. In particular,
it can follow many more spins for much longer integration time. 
We use these techniques and a model designed
specifically to emulate BCC iron and have been able to
unequivocally identify propagating spin wave modes in the
paramagnetic state, lending substantial support to Lynn's \cite{Lynn1}
experimental findings. Interestingly, spin waves are found despite
only limited magnetic SRO.

To describe the high temperature dynamics we use a classical 
Heisenberg model ${\cal H}=-(1/2) \sum_ {\bf r \neq \bf r'}
J_{\bf r ,\bf r'}{\bf S}_{\bf r} \cdot {\bf S}_{\bf r'}$, for
which the exchange interactions, $J_{\bf r ,\bf r'}$, are obtained
from first principles electronic structure calculations. For Fe
this is a reasonable approximation since the size of the
magnetic moments associated with individual Fe-sites are only
weakly dependent on the magnetic state \cite{Pindor83} and by
including interactions up to fourth nearest neighbors it is
possible to obtain a reasonably good $T_c$.

Large scale computer simulations using SD techniques to study the
dynamic properties of Heisenberg ferromagnets \cite{Landau1} and
antiferromagnets \cite{Landau2} have been quite effective, and the
direct comparison of RbMnF$_3$ SD simulations with experiments was
especially satisfying \cite{Landau2}. We have adopted these
techniques and used
$L\times L \times L$ BCC lattices with periodic boundary
conditions and $L=32$ and 40. At each lattice site, there 
is a three-dimensional classical spin of
unit length (we absorb spin moments into the definition of the
interaction parameters) and each spin has a total of $50$ interacting
neighbors. We use interaction parameters, $J_i$, for
the $T=0$ ferromagnetic state of BCC Fe calculated using the
standard formulation \cite{Liechenstein87} and the layer-KKR method
\cite{Schulthess98}. The calculated values are $J_1=36.3386$ meV, $J_2=20.6520$ meV,
$J_3=-1.625962$ meV, and $J_4=-2.39650$ meV.

In our simulations, a hybrid Monte Carlo method
was used to study the static properties and to
generate equilibrium configurations as initial states for
integrating the coupled equations of motion of SD \cite{Landau3}.
At $T_c$ and for $L=32$, the measured nonlinear relaxation time in
the equilibrating process and the linear relaxation time between
equilibrated states for the total energy and for the magnetization
\cite{Landau5} are both smaller than $500$ hybrid steps per spin.
We discarded $5000$ hybrid steps (for equilibration) and used
every $5000^{th}$ hybrid step's state as an initial
state for the SD simulations. For the $J_i$'s used here,
$T_c=919(1)K$, which is slightly smaller than the experimental
value $T_c ^{exp}=1043K$. The equilibrium magnetization $|{\bf m}|
\equiv (1/N)|\sum_{\bf r}{\bf S}_{\bf r}| \sim (1-T/T_c)^{1/3} $
in the vicinity of $T_c$ and this is in agreement with experiments.

The SD equations of motion are
\begin{equation}
\frac {d {\bf S} _{\bf r}}{dt} = {\bf H} _{eff} \times {\bf S _{\bf r}},
\end{equation}
where ${\bf H} _{eff} \equiv -\sum_ {\bf r '} J_{\bf r ,\bf r'}
\bf S _{\bf r'} $ is an effective field at site ${\bf r}$ due to
its interacting neighbors. The integration of the equations
determines the time dependence of each spin and was
carried out using an algorithm based on second-order
Suzuki-Trotter decompositions of exponential operators as
described in \cite{Landau4}. The algorithm views each spin as
undergoing Larmor precession around its effective field ${\bf H}
_{eff}$, which is itself changing with time. To deal with the fact that
we are considering four shells of interacting neighbors, the BCC
lattice is decomposed into sixteen sublattices. This
algorithm allows time steps as large as $\delta t=0.05$ (in
units of $t_0=J_1^{-1}$). Typically, the integration was carried out
to $t_{max}=20 000\delta t=1000t_0$.

The space- and time-displaced spin-spin correlation function
$C^k({\bf r} - {\bf r'},t)$ and the related dynamical structure
factor, $S^k({\bf q},\omega)$, are fundamental in the study of
spin dynamics
\cite{Lovesey} and are defined as
\begin{equation}
C^k({\bf r} - {\bf r'},t) =\langle {S_{{\bf r}}}^k(t){S_{{\bf r'}}}^k(0)\rangle-
\langle {S_{{\bf r}}}^k(t)\rangle\langle {S_{{\bf r'}}}^k(0)\rangle,
\label{eq:C}
\end{equation}
where $k=x$, $y$ or $z$ and the angle brackets $\langle \cdots
\rangle$ denote the ensemble average, and
\begin{equation}
S^k({\bf q},\omega)=\sum_{{\bf r,r'}} e^{i {\bf q}\cdot ({\bf
r}-{\bf r'})} \int_{-\infty}^{+\infty} e^{i\omega t} C^k({\bf r} -
{\bf r'},t) \frac{dt}{\sqrt{2\pi}}, \label{eq:S}
\end{equation}
where ${\bf q}$ and $\omega$ are momentum and energy ($E
\propto \omega$) transfer respectively. It is $S^k({\bf
q},\omega)$ that was probed in the neutron scattering experiments
discussed earlier.

By calculating partial spin sums `on the fly' \cite{Landau1}, it
is possible to calculate $S^k({\bf q},\omega)$ without storing a
huge amount of data associated with each spin configuration.
Because $L$ is finite, only a finite set of $q$ values are accessible:
$q=2\pi n_q/(La)$ with $n_q=\pm 1, \pm 2, \ldots, \pm L$ for the
$(q,0,0)$ and $(q,q,q)$ directions and $n_q=\pm 1, \pm 2, \ldots, \pm L/2$
for the $(q,q,0)$ direction. ($a$ is lattice constant.)
For $T\geq T_c$, the ensemble average in
Eq.\ \ref{eq:C} was performed using at least 2000 
starting configurations.
We average $S^k({\bf q},\omega)$ over
equivalent directions and this averaged structure factor is denoted
as $S({\bf q},\omega )$.

\begin{figure}
\centerline{
\includegraphics[clip,scale=0.3,angle=270]{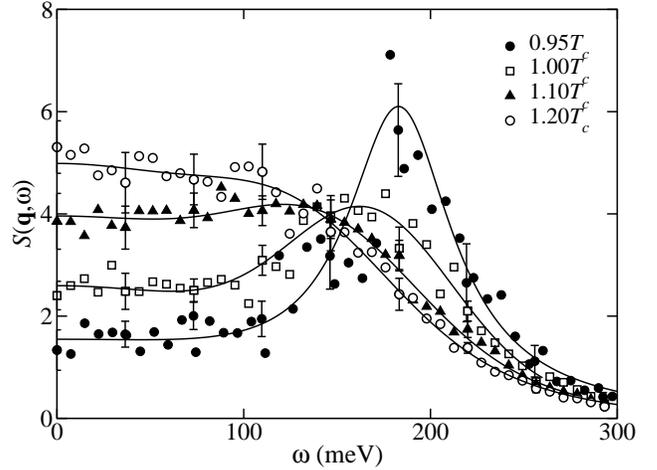}}
\caption{\label{highT100q_pi} Calculated energy dependence of
$S({\bf q},\omega )$ at ${\bf q}=\pi/a (1,0,0)$ and for
$T=0.95T_c$ (700 runs), $1.0T_c$ (2000 runs), $1.1T_c$ (2240
runs), and $1.2T_c$ (2240 runs) for $L=32$.
The solid lines are fits to the data as explained in text. 
Error bars are shown at a few typical points.}
\end{figure}

In Fig.~\ref{highT100q_pi} we show the frequency dependence of
$S({\bf q},\omega )$ obtained for four different temperatures around $T_c$.
%==================================================================
%Figure 1..........................................................
%==================================================================
%==================================================================
These, so called, constant-{$\bf q$}  scans are for ${\bf q}=\pi/a
(1,0,0)$ ($|{\bf q}|=1.09$ \AA $^{-1}$), which is half way to
Brillouin Zone boundary. At
$0.95T_c$, $S({\bf q},\omega )$ already has a 3-peak structure:
one weak central peak at zero energy and two symmetric spin wave
peaks (we only show data for $\omega \geq 0$ since the structure
factor is symmetric about $\omega = 0$). Note that the spin wave
peaks are already quite wide. As $T$ goes to $T_c$ and above, the
central peak becomes more pronounced. In addition, the spin wave
peaks shift to lower energies, broaden further and become less
obvious, however they still persist. This 3-peak structure at high
temperatures is in contrast to the 2-peak spin wave structure
found at low temperatures. In the neutron scattering from
$^{54}$Fe($12\%$Si) experiments \cite{Lynn2}, Mook and Lynn also
noticed a central peak, but could not decide whether it was
intrinsic to pure iron or a result of alloying of silicon.

In general, constant-{\bf q} scans are isotropic in the $(q,0,0)$,
$(q,q,0)$, and $(q,q,q)$ directions. For very small $|{\bf q}|$,
there is only a central peak in the scans (as is expected) and the
3-peak structure only develops for larger $|{\bf q}|$. We fit 
the 3 peaks in $S({\bf q},\omega )$
using different fitting functions and found the best results with
either a Gaussian central peak plus two Lorentzian peaks at $\pm
\omega_0$:
\begin{eqnarray}
S({\bf q},\omega )= G + L_+ + L_- , \label{1G2L}
\end{eqnarray}
or a Gaussian central peak plus two additional Gaussian peaks at $\pm \omega_0$:
\begin{eqnarray}
S({\bf q},\omega )= G + G_+ + G_- , \label{3G}
\end{eqnarray}
where  $G =I_cexp(-\omega ^2/\omega_c ^2)$, $L_\pm=I_0\omega _1
^2/((\omega \mp \omega _0)^2+\omega _1 ^2)$, and
$G_\pm=I_0exp(-(\omega \mp \omega _0)^2/\omega _1 ^2)$.  For
moderate $|{\bf q}|$ the results are fit best with Eq.~\ref{1G2L},
while Eq.\ \ref{3G} works better at larger $|{\bf q}|$.
%==================================================================
%Figure 2..........................................................
%==================================================================
\begin{figure}
\centerline{
\includegraphics[clip,scale=0.4,angle=0]{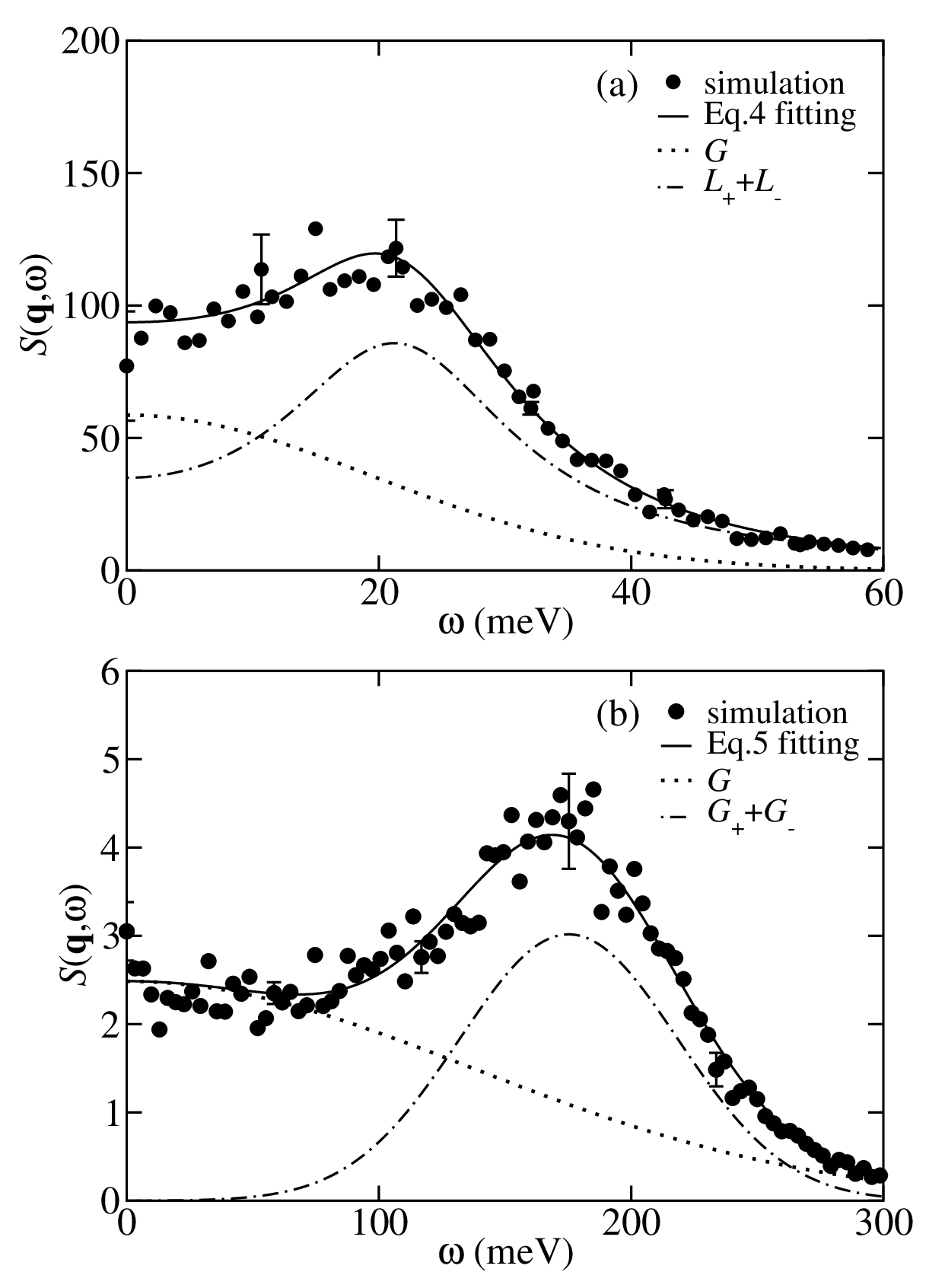}}
\caption{\label{fittingTc} Fits to $S({\bf q},\omega )$ at $T=T_c$
for two $|{\bf q}|$-points along the $(q,q,0)$ direction 
for $L=32$. (a) $|{\bf q}|=0.48$ \AA $^{-1}$ fit to
Eq.\ \ref{1G2L}, with $I_c=58.7$, $\omega_c=27.6$ meV, $I_0=80.6$,
$\omega _0=21.4$ meV, and $\omega _1=11.3$ meV; (b) $|{\bf
q}|=1.16$ \AA $^{-1}$ fit to Eq.\ \ref{3G} with $I_c=2.49$,
$\omega_c=193.3$ meV, $I_0=3.02$, $\omega _0=175.3$ meV, and
$\omega _1=61.2$ meV. The vertical scale in (b) is
much smaller than that in (a). Error bars are shown at a few
typical points.}
\end{figure}
%==================================================================
In Fig.\ \ref{fittingTc} we show, for $T=T_c$, the results of
fitting constant-${\bf q}$ scans at $|{\bf q}|=0.48$ \AA $^{-1}$ and
$|{\bf q}|=1.16$ \AA $^{-1}$ in the $(q,q,0)$ direction. 
The $|{\bf q}|=0.48$ \AA $^{-1}$ result fits well to
Eq.\ \ref{1G2L} and has $\omega _1/\omega _0<1$, i.e., the
excitation lifetime is longer than its period and thus it can be
regarded as a spin wave excitation. It should be noted that this
$|{\bf q}|$ value is very close to that ($0.47$ \AA $^{-1}$) for
which Lynn found propagating modes in contradiction to the findings of the BNL group.
At $|{\bf q}|=1.16$ \AA $^{-1}$, the structure factor has much
weaker intensity and fits best to Eq.\ \ref{3G} with a ratio $\omega
_1/\omega _0$ that is even smaller than at $|{\bf q}|=0.48$ \AA
$^{-1}$. This is illustrative of the general conclusion that
the propagating nature of the excitation modes is most pronounced
at large $|{\bf q}|$.

\begin{figure}
\centerline{
\includegraphics[clip,scale=0.35,angle=0]{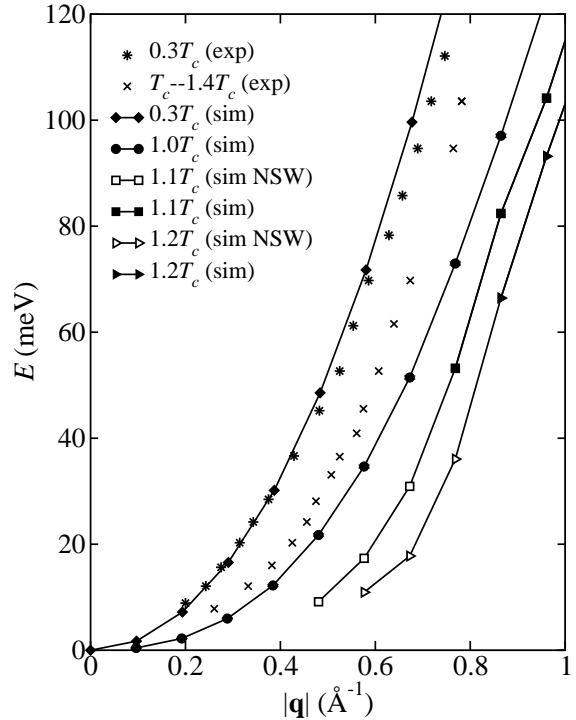}}
\caption{\label{dispersion} Comparison of dispersion curves
obtained in our simulations (sim) with Lynn's experimental (exp)
(Ref. \cite{Lynn1}) results for the $(q,q,0)$ direction. Open
symbols indicate excitations with mixed nature and are not due to
spin waves (NSW).}
\end{figure}

Figure~\ref{dispersion} shows the dispersion relations obtained by
plotting the peak positions, $\omega_0$, determined from the fits
to $S({\bf q},\omega )$ along the $(q,q,0)$
direction. Calculated dispersion curves are shown at several
temperatures in the ferromagnetic and paramagnetic phases together
with the experimental results of Lynn \cite{Lynn1}.
%==================================================================
%Figure 3..........................................................
%==================================================================
%==================================================================
To estimate errors, we fitted each constant-${\bf q}$ scan several
times by cutting off the tail at slightly different $\omega_{max}$
to get an average $\omega_0$; these error bars are found to be no
larger than symbols. In this figure, filled symbols
indicate modes that are clearly propagating ($\omega _1/\omega
_0<1$) while open symbols indicate that, even though there are
peaks at $\omega _0 \neq 0$, the peaks have widths $\omega _1 >
\omega _0$. The calculated result for $T=0.3T_c$ is very close to
that from the experiments and propagating modes exist for very
small $|{\bf q}|$. For $T \geq T_c$, our curves lie below the
experiments's and soften with increasing temperatures, a property
not seen in the experiments. One possibility deserving of further
study is that our use of temperature and configuration independent
exchange interactions, in particular those appropriate to the
$T=0$ ferromagnetic state, breaks down at high temperatures when
the spin moments are highly non-collinear.

\begin{figure}
\centerline{
\includegraphics[clip,scale=0.34,angle=270]{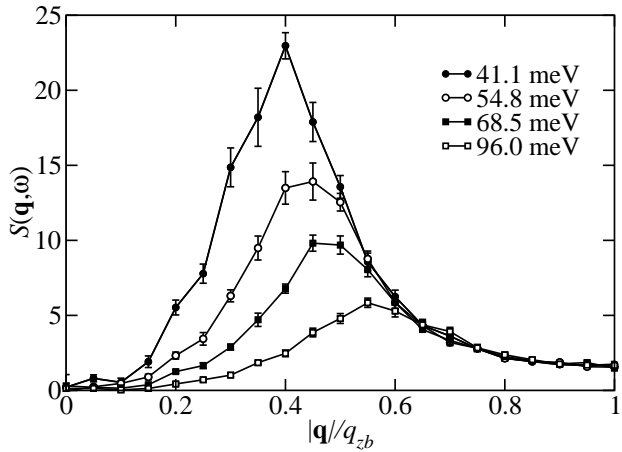}}
\caption{\label{constE40} $T=1.1T_c$ constant-$E$ scans along
$(q,q,0)$ direction for $E=41.1$ meV, $54.8$ meV, $68.5$ meV, and
$96.0$ meV with $L=40$. Brillouin Zone boundary $q_{zb}=1.55$ \AA $^{-1}$ in the direction.
}
\end{figure}

In our simulations we have equal access to constant-${\bf q}$
scans and constant-$E$ scans; however, this is not the case in
neutron scattering experiments. Because the dispersion curves of
Fe are generally very steep, experimentalists usually perform
constant-$E$ scans. In Fig.~\ref{constE40} we show constant-$E$
scans for several $E$ values at
$T=1.1T_c$ based on simulations. Clearly, the
constant-$E$ scans have two peaks (symmetric about $|{\bf q}|=0$)
that become smaller and wider and shift to higher $|{\bf q}|$ as
$E$ increases. Peaks in constant-$E$ scans strongly suggest that
SRO persists above $T_c$ \cite{Korenman1}.
%==================================================================
%Figure 4..........................................................
%==================================================================
%==================================================================

The degree of magnetic SRO can be obtained directly from the
behavior of static correlation function $C^k({\bf r} - {\bf
r'},0)$ (i.e. Eq.~\ref{eq:C} with $t=0$), which can be calculated
from the Monte Carlo configurations alone. For $T=1.1T_c$ we find
a correlation length of approximately $2a$ ($\sim 6$
neighbor shells), indicative of only limited SRO. Thus, in general,
extensive SRO is not required to support spin waves. Moreover, inspection of
Fig.~\ref{dispersion} for $T=1.1T_c$ shows that the point $q \gtrsim
0.77$\AA$^{-1}$, at which these peaks first correspond to
propagating modes, is when their wavelength ($\lambda \sim
2a$) first becomes the order of the static correlation length.

In summary, our SD simulations clearly point to the existence of
spin waves in the paramagnetic state of BCC Fe and support the
original conclusions of Lynn. Their signature is seen as spin wave
peaks in dynamical structure factor in constant-${\bf q}$ and
constant-$E$ scans. Detailed analysis of the constant-${\bf q}$
scans shows that the propagating nature of these excitations is
clearest at large $|{\bf q}|$, in agreement with experiment. This
is also consistent with the requirement that their wavelength be
the order of, or shorter than, the static correlation length.
While the inclusion of four shells of first-principles-determined
interactions into the Heisenberg model makes our results
specifically relate to BCC Fe, we have also found spin waves in a
Heisenberg model containing only nearest neighbor interactions. In
addition to elucidating the longstanding controversy regarding
the existence of spin waves above $T_c$, these simulations also point
to the important role that inelastic neutron scattering studies of
the paramagnetic state can have in understanding the nature of
magnetic excitations, particularly when coupled with
state-of-the-art SD simulations.

\begin{acknowledgments}
We thank Shan-Ho Tsai, H. K. Lee, V. P. Antropov, and K. Binder
for informative discussions. Computations were performed at
ORNL-CCS (www.ccs.ornl.gov) and NERSC (www.nersc.gov) using
elements of the $\Psi$-\textit{Mag} toolset which may be obtained
at http://mri-fre.ornl.gov/psimag. Work supported by Computational
Materials Science Network sponsored by DOE BES-DMSE (XT), BES-DMSE
(GMS), NSF Grant No. DMR-0341874 (XT and DPL) and DARPA (XT and
TCS) under contract No. DE-AC05-00OR22725 with UT-Battelle LLC.
\end{acknowledgments}


\begin{thebibliography}{24}
\expandafter\ifx\csname natexlab\endcsname\relax\def\natexlab#1{#1}\fi
\expandafter\ifx\csname bibnamefont\endcsname\relax
  \def\bibnamefont#1{#1}\fi
\expandafter\ifx\csname bibfnamefont\endcsname\relax
  \def\bibfnamefont#1{#1}\fi
\expandafter\ifx\csname citenamefont\endcsname\relax
  \def\citenamefont#1{#1}\fi
\expandafter\ifx\csname url\endcsname\relax
  \def\url#1{\texttt{#1}}\fi
\expandafter\ifx\csname urlprefix\endcsname\relax\def\urlprefix{URL }\fi
\providecommand{\bibinfo}[2]{#2}
\providecommand{\eprint}[2][]{\url{#2}}

\bibitem[{\citenamefont{Collins et~al.}(1969)\citenamefont{Collins, Minkiewicz,
  Nathans, Passell, and Shirane}}]{Collins}
\bibinfo{author}{\bibfnamefont{M.~F.} \bibnamefont{Collins}},
  \bibinfo{author}{\bibfnamefont{V.~J.} \bibnamefont{Minkiewicz}},
  \bibinfo{author}{\bibfnamefont{R.}~\bibnamefont{Nathans}},
  \bibinfo{author}{\bibfnamefont{L.}~\bibnamefont{Passell}}, \bibnamefont{and}
  \bibinfo{author}{\bibfnamefont{G.}~\bibnamefont{Shirane}},
  \bibinfo{journal}{Phys. Rev.} \textbf{\bibinfo{volume}{179}},
  \bibinfo{pages}{417} (\bibinfo{year}{1969}).

\bibitem[{\citenamefont{Lynn}(1975)}]{Lynn1}
\bibinfo{author}{\bibfnamefont{J.~W.} \bibnamefont{Lynn}},
  \bibinfo{journal}{Phys. Rev. B} \textbf{\bibinfo{volume}{11}},
  \bibinfo{pages}{2624} (\bibinfo{year}{1975}).

\bibitem[{\citenamefont{Wicksted
  et~al.}(1984{\natexlab{a}})\citenamefont{Wicksted, Shirane, and
  Steinsvoll}}]{Shirane1}
\bibinfo{author}{\bibfnamefont{J.~P.} \bibnamefont{Wicksted}},
  \bibinfo{author}{\bibfnamefont{G.}~\bibnamefont{Shirane}}, \bibnamefont{and}
  \bibinfo{author}{\bibfnamefont{O.}~\bibnamefont{Steinsvoll}},
  \bibinfo{journal}{Phys. Rev. B} \textbf{\bibinfo{volume}{29}},
  \bibinfo{pages}{R488} (\bibinfo{year}{1984}{\natexlab{a}});
\bibinfo{author}{\bibfnamefont{G.}~\bibnamefont{Shirane}},
  \bibinfo{author}{\bibfnamefont{O.}~\bibnamefont{Steinsvoll}},
  \bibinfo{author}{\bibfnamefont{Y.~J.} \bibnamefont{Uemura}},
  \bibnamefont{and} \bibinfo{author}{\bibfnamefont{J.}~\bibnamefont{Wicksted}},
  \bibinfo{journal}{J. Appl. Phys.} \textbf{\bibinfo{volume}{55}},
  \bibinfo{pages}{1887} (\bibinfo{year}{1984});
\bibinfo{author}{\bibfnamefont{J.~P.} \bibnamefont{Wicksted}},
  \bibinfo{author}{\bibfnamefont{P.}~\bibnamefont{Boni}}, \bibnamefont{and}
  \bibinfo{author}{\bibfnamefont{G.}~\bibnamefont{Shirane}},
  \bibinfo{journal}{Phys. Rev. B} \textbf{\bibinfo{volume}{30}},
  \bibinfo{pages}{3655} (\bibinfo{year}{1984}{\natexlab{b}}).

\bibitem[{\citenamefont{Mook and Lynn}(1985)}]{Lynn2}
\bibinfo{author}{\bibfnamefont{H.~A.} \bibnamefont{Mook}} \bibnamefont{and}
  \bibinfo{author}{\bibfnamefont{J.~W.} \bibnamefont{Lynn}},
  \bibinfo{journal}{J. Appl. Phys.} \textbf{\bibinfo{volume}{57}},
  \bibinfo{pages}{3006} (\bibinfo{year}{1985}).

\bibitem[{\citenamefont{Haines et~al.}(1985{\natexlab{a}})\citenamefont{Haines,
  Heine, and Ziegler}}]{Haines1}
\bibinfo{author}{\bibfnamefont{E.~M.} \bibnamefont{Haines}},
  \bibinfo{author}{\bibfnamefont{V.}~\bibnamefont{Heine}}, \bibnamefont{and}
  \bibinfo{author}{\bibfnamefont{A.}~\bibnamefont{Ziegler}},
  \bibinfo{journal}{J. Phys. F: Metal Phys.} \textbf{\bibinfo{volume}{15}},
  \bibinfo{pages}{661} (\bibinfo{year}{1985}{\natexlab{a}});
\bibinfo{author}{\bibfnamefont{E.~M.} \bibnamefont{Haines}},
  \bibinfo{author}{\bibfnamefont{R.}~\bibnamefont{Clauberg}}, \bibnamefont{and}
  \bibinfo{author}{\bibfnamefont{R.}~\bibnamefont{Feder}},
  \bibinfo{journal}{Phys. Rev. Lett.} \textbf{\bibinfo{volume}{54}},
  \bibinfo{pages}{932} (\bibinfo{year}{1985}{\natexlab{b}}).

\bibitem[{\citenamefont{Kisker et~al.}(1985)\citenamefont{Kisker, Clauberg, and
  Gudat}}]{Kisker}
\bibinfo{author}{\bibfnamefont{E.}~\bibnamefont{Kisker}},
  \bibinfo{author}{\bibfnamefont{R.}~\bibnamefont{Clauberg}}, \bibnamefont{and}
  \bibinfo{author}{\bibfnamefont{W.}~\bibnamefont{Gudat}}, \bibinfo{journal}{Z.
  Phys. B} \textbf{\bibinfo{volume}{61}}, \bibinfo{pages}{453}
  (\bibinfo{year}{1985}).

\bibitem[{\citenamefont{Korenman et~al.}(1977)\citenamefont{Korenman, Murray,
  and Prange}}]{Korenman1}
\bibinfo{author}{\bibfnamefont{V.}~\bibnamefont{Korenman}},
  \bibinfo{author}{\bibfnamefont{J.~L.} \bibnamefont{Murray}},
  \bibnamefont{and} \bibinfo{author}{\bibfnamefont{R.~E.}
  \bibnamefont{Prange}}, \bibinfo{journal}{Phys. Rev. B}
  \textbf{\bibinfo{volume}{16}}, \bibinfo{pages}{4032} (\bibinfo{year}{1977});
\bibinfo{author}{\bibfnamefont{R.~E.}~\bibnamefont{Prange}} \bibnamefont{and}
  \bibinfo{author}{\bibfnamefont{V.}~\bibnamefont{Korenman}},
  \bibinfo{journal}{Phys. Rev. B} \textbf{\bibinfo{volume}{19}},
  \bibinfo{pages}{4691} (\bibinfo{year}{1978}).

\bibitem[{\citenamefont{Capellmann}(1979{\natexlab{a}})}]{Capellmann1}
\bibinfo{author}{\bibfnamefont{H.}~\bibnamefont{Capellmann}},
  \bibinfo{journal}{Solid State Commun.} \textbf{\bibinfo{volume}{30}},
  \bibinfo{pages}{7} (\bibinfo{year}{1979}{\natexlab{a}});
  \bibinfo{journal}{Z. Phys. B} \textbf{\bibinfo{volume}{35}},
  \bibinfo{pages}{269} (\bibinfo{year}{1979}{\natexlab{b}}).

\bibitem[{\citenamefont{Heine and Joynt}(1988)}]{Heine}
\bibinfo{author}{\bibfnamefont{V.}~\bibnamefont{Heine}} \bibnamefont{and}
  \bibinfo{author}{\bibfnamefont{R.}~\bibnamefont{Joynt}},
  \bibinfo{journal}{Europhys. Lett.} \textbf{\bibinfo{volume}{5}},
  \bibinfo{pages}{81} (\bibinfo{year}{1988}).

\bibitem[{\citenamefont{Hubbard}(1979)}]{Hubbard1}
\bibinfo{author}{\bibfnamefont{J.}~\bibnamefont{Hubbard}},
  \bibinfo{journal}{Phys. Rev. B} \textbf{\bibinfo{volume}{20}},
  \bibinfo{pages}{4584} (\bibinfo{year}{1979});
  \bibinfo{journal}{Phys. Rev. B} \textbf{\bibinfo{volume}{23}},
  \bibinfo{pages}{5974} (\bibinfo{year}{1981}).

\bibitem[{\citenamefont{Moriya}(1991)}]{Moriya}
\bibinfo{author}{\bibfnamefont{T.}~\bibnamefont{Moriya}}, \bibinfo{journal}{J.
  Magn. Magn. Mat.} \textbf{\bibinfo{volume}{100}}, \bibinfo{pages}{261}
  (\bibinfo{year}{1991}).

\bibitem[{\citenamefont{Shastry et~al.}(1981)\citenamefont{Shastry, Edwards,
  and Young}}]{Shastry1}
\bibinfo{author}{\bibfnamefont{B.~S.} \bibnamefont{Shastry}},
  \bibinfo{author}{\bibfnamefont{D.~M.} \bibnamefont{Edwards}},
  \bibnamefont{and} \bibinfo{author}{\bibfnamefont{A.~P.} \bibnamefont{Young}},
  \bibinfo{journal}{J. Phys. C} \textbf{\bibinfo{volume}{14}},
  \bibinfo{pages}{L665} (\bibinfo{year}{1981}).

\bibitem[{\citenamefont{Shastry}(1984)}]{Shastry2}
\bibinfo{author}{\bibfnamefont{B.~S.} \bibnamefont{Shastry}},
  \bibinfo{journal}{Phys. Rev. Lett.} \textbf{\bibinfo{volume}{53}},
  \bibinfo{pages}{1104} (\bibinfo{year}{1984}).

  \bibitem[{\citenamefont{Pindor}(1983)}]{Pindor83}
\bibinfo{author}{\bibfnamefont{A.~J.} \bibnamefont{Pindor}},
\bibinfo{author}{\bibfnamefont{J.} \bibnamefont{Staunton}},
\bibinfo{author}{\bibfnamefont{G.~M.} \bibnamefont{Stocks}}, \bibnamefont{and}
\bibinfo{author}{\bibfnamefont{H.} \bibnamefont{Winter}},
  \bibinfo{journal}{J. Phys. F} \textbf{\bibinfo{volume}{13}},
  \bibinfo{pages}{979} (\bibinfo{year}{1983}).

\bibitem[{\citenamefont{Chen and Landau}(1994)}]{Landau1}
\bibinfo{author}{\bibfnamefont{K.}~\bibnamefont{Chen}} \bibnamefont{and}
  \bibinfo{author}{\bibfnamefont{D.~P.} \bibnamefont{Landau}},
  \bibinfo{journal}{Phys. Rev. B} \textbf{\bibinfo{volume}{49}},
  \bibinfo{pages}{3266} (\bibinfo{year}{1994}).

\bibitem[{\citenamefont{Tsai et~al.}(2000)\citenamefont{Tsai, Bunker, and
  Landau}}]{Landau2}
\bibinfo{author}{\bibfnamefont{S.-H.} \bibnamefont{Tsai}},
  \bibinfo{author}{\bibfnamefont{A.}~\bibnamefont{Bunker}}, \bibnamefont{and}
  \bibinfo{author}{\bibfnamefont{D.~P.} \bibnamefont{Landau}},
  \bibinfo{journal}{Phys. Rev. B} \textbf{\bibinfo{volume}{61}},
  \bibinfo{pages}{333} (\bibinfo{year}{2000}).

  \bibitem[{\citenamefont{Liechenstein}(1987)}]{Liechenstein87}
\bibinfo{author}{\bibfnamefont{A.~I.} \bibnamefont{Liechenstein}},
\bibinfo{author}{\bibfnamefont{M.~I.} \bibnamefont{Katsnelson}},
\bibinfo{author}{\bibfnamefont{V.~P.} \bibnamefont{Antropov}}, \bibnamefont{and}
\bibinfo{author}{\bibfnamefont{V.~A.} \bibnamefont{Gubanov}},
  \bibinfo{journal}{J. Mag. Mag. Mater.} \textbf{\bibinfo{volume}{67}},
  \bibinfo{pages}{65} (\bibinfo{year}{1987}).

  \bibitem[{\citenamefont{Schulthess}(1998)}]{Schulthess98}
  \bibinfo{author}{\bibfnamefont{T.~C.}~\bibnamefont{Schulthess}} \bibnamefont{and}
  \bibinfo{author}{\bibfnamefont{W.~H.}~\bibnamefont{Butler}},
  \bibinfo{journal}{J. App. Phys.} \textbf{\bibinfo{volume}{83}},
  \bibinfo{pages}{7225} (\bibinfo{year}{1998}).

\bibitem[{\citenamefont{Peczak and Landau}(1993)}]{Landau3}
\bibinfo{author}{\bibfnamefont{P.}~\bibnamefont{Peczak}} \bibnamefont{and}
  \bibinfo{author}{\bibfnamefont{D.~P.} \bibnamefont{Landau}},
  \bibinfo{journal}{Phys. Rev. B} \textbf{\bibinfo{volume}{47}},
  \bibinfo{pages}{14260} (\bibinfo{year}{1993}).

\bibitem[{\citenamefont{Landau and Binder}(2000)}]{Landau5}
\bibinfo{author}{\bibfnamefont{D.~P.} \bibnamefont{Landau}} \bibnamefont{and}
  \bibinfo{author}{\bibfnamefont{K.}~\bibnamefont{Binder}},
  \emph{\bibinfo{title}{A Guide to Monte Carlo Simulations in Statistical
  Physics}} (\bibinfo{publisher}{Cambridge University Press},
  \bibinfo{address}{Cambridge}, \bibinfo{year}{2000}).

\bibitem[{\citenamefont{Krech et~al.}(1998)\citenamefont{Krech, Bunker, and
  Landau}}]{Landau4}
\bibinfo{author}{\bibfnamefont{M.}~\bibnamefont{Krech}},
  \bibinfo{author}{\bibfnamefont{A.}~\bibnamefont{Bunker}}, \bibnamefont{and}
  \bibinfo{author}{\bibfnamefont{D.~P.} \bibnamefont{Landau}},
  \bibinfo{journal}{Comput. Phys. Commun.} \textbf{\bibinfo{volume}{111}},
  \bibinfo{pages}{1} (\bibinfo{year}{1998}).

\bibitem[{\citenamefont{Lovesey}(1984)}]{Lovesey}
\bibinfo{author}{\bibfnamefont{S.~W.} \bibnamefont{Lovesey}},
  \emph{\bibinfo{title}{Theory of neutron scattering from condensed matter}}
  (\bibinfo{publisher}{Clarendon}, \bibinfo{address}{Oxford},
  \bibinfo{year}{1984}).

\end{thebibliography}
\end{document}